\documentclass[12pt]{article}
\def\vep{\varepsilon}
\def\al{\alpha}

\def\beq{\begin{equation}}
\def\eeq{\end{equation}}
\def\pd{\partial}
\def\bea{\begin{eqnarray}}
\def\eea{\end{eqnarray}}

\begin{document}
\begin{center}
\Large{
{\bf Extended BRST quantization in general coordinates}}

\vspace{0.5cm}
{\large B. Geyer}$^{a}$\footnote{e-mail: geyer@itp.uni-leipzig.de},
{\large P. Lavrov}$^{b}$\footnote{e-mail:lavrov@tspu.edu.ru}and
{\large A. Nersessian}$^{c}$
\end{center}
\vspace{0.5cm}
{\it $^{a}$ Center of Theoretical Studies, Leipzig University,
Augustusplatz 10/11, D-04109 Leipzig, Germany\\
 $^{b}$ Tomsk State Pedagogical University,
634041 Tomsk, Russia\\
$^{c}$ Yerevan State University, A. Manoogian St., 3,
Yerevan, 375025, Armenia}

\begin{abstract}
 We propose an extended BRST invariant Lagrangian quantization scheme
 of general gauge theories  based on explicit realization
 of "modified triplectic algebra" in general coordinates.
 All the known Lagrangian  quantization schemes based
 on the extended BRST symmetry are obtained by specifying the (free)
 parameters of that method.
\end{abstract}

\section{Introduction}

Modern covariant quantization methods of general gauge
theories are based on the principle of the BRST~\cite{BV} or
the extended BRST~\cite{BLT,BM,BMS,BM1,ND,GGL} invariance.
Consideration of these methods in general coordinates seems to be a very
important task for understanding the geometrical sense of main objects,
like (extended) antibracket(s), generating second-order operator(s)
and so on, underlying these schemes.

Investigations of the well-known Batalin-Vilkovisky (BV)
method~\cite{BV} in general coordinates were initiated by
E.~Witten~\cite{W} and A.~Schwarz~\cite{S}. It was found that
the geometry of the BV formalism is the geometry of a superspace
equipped with an {\it odd} symplectic structure $\omega$ and a density
function $\rho$~\cite{bvgeom}.  It will be shown below that
the realization of such a programme for the quantization methods, based
on the extended BRST symmetry, seems to be more complicated and, in
addition, requires to use not only an {\it even} Poisson structure and
a density function but also a flat symplectic connection and a
symmetric structure (like a metric tensor).

\section{Quantization methods in Darboux coordinates}

Let us present the basic ingredients of extended BRST invariant Lagrangian
quantization schemes, thereby following the original papers
\cite{BLT,BM,BMS,GGL}. Among them the first was the so-called $Sp(2)$ -
covariant quantization proposed in papers~\cite{BLT}. The formulation of
that method includes the introduction of a configuration space of fields
$\phi^A$ with Grassmann parity $\epsilon(\phi^A)=\epsilon_A$
which contains the fields of the initial classical theory, ghost and
antighost fields, auxiliary fields and so on. For the purpose of
the present investigation the explicit structure of the configuration
space is not important and will not be discussed furthermore. To each
field $\phi^A$ it is necessary to introduce three kinds of antifields,
namely, a $Sp(2)$--doublet $\phi^*_{Aa}$
($\epsilon(\phi^*_{Aa})=\epsilon_A+1$) where $a=1,2$ is the index of the
group $Sp(2)$, and a singlet $\bar{\phi}_A$
($\epsilon(\bar{\phi}_A)=\epsilon_A$). The antifields $\phi^*_{Aa}$ can be
considered as sources of BRST and antiBRST transformations while the
antifields $\bar{\phi}_A$ are sources of the mixed BRST-antiBRST
transformations. Furthermore, the set of auxiliary $Sp(2)$--doublet
fields $\pi^{Aa}$ ($\epsilon(\pi^{Aa})=\epsilon_A+1$) and singlet field
$\lambda^A$ ($\epsilon(\lambda^A)=\epsilon_A$) is introduced.  On the
superspace of fields $\phi^A$ and antifields  $\phi^*_{Aa}$ the
extended antibrackets $(\;,\;)^a$ are defined for any two functionals
$F$ and $G$ by the rule
\begin{equation}
\label{1} (F,G)^a=
\frac{\pd_r F}{\pd\phi^A} \frac{\pd G}{\pd\phi^*_{Aa}} - (F\leftrightarrow G)
(-1)^{(\epsilon(F)+1)(\epsilon(G)+1)}.
\end{equation}
Usually,
all derivatives are considered as left derivatives; right derivatives
are labelled by a special mark "r". Among the properties of the
extended antibrackets there exist non-trivial relations,
\begin{equation} \label{Ja}
((F,G)^{\{a},H)^{b\}}(-1)^{(\epsilon(F)+1)(\epsilon(H)+1)}+cycle(F,G,H)
\equiv 0,
\end{equation}
which can be considered as generalized Jacobi identities. Here and
below curly brackets denote symmetrization with respect to  the
enclosed $Sp(2)$ indices $a$ and $b$
$F^{\{a}G^{b\}}\equiv F^aG^b+F^bG^a$. Then, the generating operators
$\Delta^a$ and $V^a$ are introduced,
\bea
\Delta^a=(-1)^{\epsilon_A}\frac{\pd}{\pd\phi^a}
\frac{\pd}{\pd\phi^*_{Aa}},\quad
V^a=\epsilon^{ab}\phi^*_{Ab}\frac{\pd}{\pd\bar{\phi_A}},
\label{spV}
\eea
where $\epsilon^{ab}$ is the antisymmetric tensor of the $Sp(2)$ group.
The operators $\Delta^a$ and $V^a$ form the algebra of nilpotent and
anticommuting operators,
\begin{equation}
\label{spalg}
\Delta^{\{a}\Delta^{b\}}=0, \quad V^{\{a}V^{b\}}=0,\quad
\Delta^{\{a}V^{b\}} +V^{\{a}\Delta^{b\}}=0.
\end{equation}

The vacuum functional within the $Sp(2)$ approach is constructed as
the following functional integral,
\begin{equation}
\label{Z}
Z=\int d\phi d\phi^* d\bar\phi d\pi d\lambda
\exp\{(i/\hbar)[S+X+S_0]\}, \end{equation}
where the bosonic functional
$S=S(\phi,\phi^*,\bar\phi)$, the extended action, satifies the quantum
master equations,
\bea
\label{qme}
\frac{1}{2}(S,S)^a+V^aS=i\hbar\Delta^a S \leftrightarrow (\Delta^a
+(i/\hbar)V^a)\exp\{(i/\hbar)S\} =0,
\eea
and the boundary condition
$S|_{\phi^*=\bar\phi=\hbar=0}=S_{cl}$, with $S_{cl}$ being the initial
classical action.  In (\ref{qme}) the gauge fixing functional,
$X=X(\phi,\pi,\bar\phi;\lambda)$, and the $Sp(2)$ scalar functional of
special form, $S_0=S_0(\phi^*,\pi)$, are introduced as follows:
\beq
\label{S0}
X=\left(\bar\phi_A-\frac{\partial G}{\pd\phi^A}\right)\lambda^A -
\frac 12\vep_{ab}\pi^{Aa}\frac{\pd^2G}{\pd\phi^A\pd\phi^B}
\pi^{Bb},\quad
S_0=\phi^*_{Aa}\pi^{Aa};
\eeq
thereby the functional
$G=G(\phi)$ defines a specific choice of the gauge.  The vacuum
functional (\ref{Z}) is invariant under the extended BRST
transformations \bea \nonumber \delta\phi^A=\pi^{Aa}\mu_a, \;
\delta\phi^*_{Aa}=\mu_a\frac{\pd S}{\pd\phi^A},\;
\delta\bar{\phi}_A=\epsilon^{ab}\mu_a\phi^*_{Ab},\;
\delta\pi^{Aa}=-\epsilon^{ab}\lambda^A\mu_b, \;
\delta\lambda^A=0,
\eea
with a $Sp(2)$-doublet of constant Grassmann parameters, $\mu_a$,
and it is gauge independent, $Z_{G+\delta G}=Z_G$.

Another version of the  $Sp(2)$ formalism, suggested in~\cite{BM},
considers the auxiliary fields $\pi^{aA}$ as canonically conjugate with
respect to the fields $\bar\phi_A$ with the following re-definitions
of the extended antibrackets,
\bea
\label{trantibr}
(F,G)^a=\frac{\pd_r F}{\pd\phi^A}
\frac{\pd G}{\pd\phi^*_{Aa}} + \epsilon^{ab}\frac{\pd_r F}{\pd\pi^{Ab}}
\frac{\pd G}{\pd\bar\phi_A} -
(F\leftrightarrow G) (-1)^{(\epsilon(F)+1)(\epsilon(G)+1)},
\eea
and generating operators $\Delta^a$ and $V^a$,
\bea
\label{trD}
\Delta^a&=&
(-1)^{\epsilon_A}\frac{\pd}{\pd\phi^a}
\frac{\pd}{\pd\phi^*_{Aa}} +(-1)^{\epsilon_A+1}\epsilon^{ab}
 \frac{\pd}{\pd\pi^{Ab}}\frac{\pd}{\pd\bar\phi_A},\\
V^a&=&
\frac{1}{2}\Big(\epsilon^{ab}\phi^*_{Ab}\frac{\pd}{\pd\bar{\phi_A}}+
(-1)^{\epsilon_A+1}\pi^{Aa}\frac{\pd}{\pd\phi^A}\Big).
\label{trV}
\eea

The operators $\Delta^a$ (\ref{trD}) and $V^a$ (\ref{trV}) form the
following  (tripletic) algebra,
\begin{equation}
\label{tralg}
\Delta^{\{a}\Delta^{b\}}=0, \quad V^{\{a}V^{b\}}=0,\quad
\Delta^aV^b +V^a\Delta^b=0.
\end{equation}
Notice that by the explicit realization of the vector fields $V^a$
in the form (\ref{trV})
the last relations in (\ref{tralg}) in comparison with (\ref{spalg})
hold without symmetrization of the indices $a$ and $b$.

The vacuum functional is constructed in the following form,
\begin{equation}
\label{trZ}
Z=\int dz d\lambda
\exp\{(i/\hbar)[S+X]\},
\end{equation}
where the bosonic functional $S=S(z)$ ($z=(\phi,\phi^*,\pi,\bar\phi)$)
satifies equations like (\ref{qme}) and the gauge fixing functional
$X=X(z,\lambda)$
is required to satisfy the following master equations:
$(\Delta^a -(i/\hbar)V^a)\exp\{(i/\hbar)X\} =0$.
The vacuum functional (\ref{trZ}) is invariant under the extended BRST
transformations
$\delta\Gamma =(\Gamma,X-S)^a\mu_a+2\mu_aV^a\Gamma$
($\Gamma =(z,\lambda)$),
and it is gauge independent, $Z_{X+\delta X}=Z_X$.

Let us notice that
due to the special structure of the operators $V^a$, (\ref{trV}),
it is not possible, in contrast to all
previously known schemes of Lagrangian quantization, to consider the
initial classical action as a natural boundary condition to the solution
of the quantum master equations, since
$(1/2)(S_{cl},S_{cl})^a+V^aS_{cl}-i\hbar\Delta^a S_{cl}\neq 0$.
This was the main reason
in Ref.~\cite{GGL} to reformulate the original triplectic scheme by a
modified one.

Remaining in the same configuration space
of fields and antifields $z=(\phi,\phi^*,\pi,{\bar\phi})$,
it was
proposed to change from the beginning the triplectic algebra
and the generating master equations by introducing an additional set of
$Sp(2)$ doublets of operators $U^a$, $\epsilon(U^a)=1$, by requiring
\begin{eqnarray}
\label{malg}
&&
\Delta^{\{a}\Delta^{b\}} =0, \quad\qquad V^{\{a}V^{b\}} =0,
\quad\qquad\Delta^{\{a} V^{b\}} + V^{\{a}\Delta^{b\}} = 0,
\\
%\end{eqnarray}
%\begin{eqnarray}
\label{mal}
&&
U^{\{a}U^{b\}}=0,\quad \Delta^{\{a} U^{b\}} + U^{\{a}\Delta^{b\}} = 0,
\quad U^{\{a}V^{b\}} +V^{\{b}U^{a\}}=0.
\end{eqnarray}
This algebra can be considered as an extension of the triplectic algebra
(\ref{tralg}) and is refered to as the
"modified triplectic algebra".  An explicit realization of the
operators may be given by $\Delta^a$ and $V^a$ as introduced in
Eq.~(\ref{trD}) and the second of Eqs.~(\ref{spV}), respectively,
and $U^a$ in the
form
\bea
U^a=(-1)^{\epsilon_A+1}\pi^{Aa}\frac{\pd}{\pd\phi^A}.
\eea

The vacuum functional within this approach formally coincides with
Eq.~(\ref{Z})
when $S=S(z)$ and
$X=X(z,\lambda)$ satisfy the quantum master equations
$(\Delta^a + (i/\hbar)
V^a){\rm exp}\{(i/\hbar) S\} =0$ and $(\Delta^a -
(i/\hbar) U^a){\rm exp}\{(i/\hbar) X \}=0$, respectively,
and the functional $S_0$ was defined in (\ref{S0}).
Now one can use the standard boundary condition for $S$
in the form of an initial classical action $S_{cl}$
when $\hbar = \phi^* = \bar\phi = 0$.

This vacuum functional is also invariant under extended BRST
transformations
$\delta\Gamma =(\Gamma,X-S)^a\mu_a+\mu_a(V^a+U^a)\Gamma$
and it is gauge independent.

\section{(Modified) triplectic algebra in general coordinates}

Now, we consider an explicit realization of the (modified) triplectic
algebra. Our starting point is an even superspace ${\cal M}_0$ described
by local coordinates $x^i$
($\epsilon(x^i)=\epsilon_i$), which in Darboux coordinates will be
identified as $x^i=(\phi^A,\bar{\phi}_A)$. The superspace
${\cal M}_0$ is equipped with a Poisson structure $\omega^{ij}$
($\epsilon(\omega^{ij})=\epsilon_i+\epsilon_j$)
with the following properties
$\omega^{ij}=-(-1)^{\epsilon_i\epsilon_j}\omega^{ji}$ and
$(-1)^{\epsilon_i\epsilon_k}
\omega^{kl}\pd_l\omega^{ij}\;+
\quad {\rm cycl.perm}\;(ijk)=0$.

In the following, for the sake of simplicity of all expressions and
relations, we
restrict ourselves to the case when all coordinates $x^i$ are even,
$\epsilon_i=0$, although in the context of quantum field theory among
$x^i$ also odd variables (ghost fields) always are present.

${\cal M}_0$ is also equipped with a symmetric connection
$\Gamma^k_{ij}=\Gamma^k_{ji}$ (or a covariant derivative $\nabla_i$)
respecting the Poisson structure
$\pd_i\omega^{kj}+ \omega^{kl}\Gamma^j_{li}-\omega^{kj}\Gamma^k_{li}=0$,
as well as a scalar (density) function $\rho=\rho(x)$.
When the Poisson structure is nondegenerate $\Gamma^k_{ij}$ coincides,
in fact, with the Fedosov connection~\cite{F} and
$({\cal M}_0,\omega,\nabla)$ is referred to as the Fedosov
space~\cite{fm}.  The curvature tensor $R^l_{\;ijk}$ of this (Fedosov)
connection has the standard form and properties:
\beq \label{R}
R^l_{\;ijk}=
\pd_j\Gamma^l_{ki}-
\pd_k\Gamma^l_{ij} +
\Gamma^m_{ki}\Gamma^l_{jm}-\Gamma^m_{ij}\Gamma^l_{km},\quad
R^l_{\;ijk}= -R^l_{\;ikj}.
\eeq

Then we consider the superspace ${\cal M}$ parametrized by the
coordinates  $z^{\mu} =(x^i,\;\theta_{ia})$,
($\epsilon(\theta_{ia})=\epsilon_i+1$),
where $\theta_{ia}$ ("antifields") are
transformed as $\pd_i=\partial/\partial x^i$
under reparametrizations of  ${\cal M}_0$. Notice that the variables
$\theta_{ia}$ in the Darboux coordinates can be identified as
$\theta_{ia}=(\phi^*_{Aa},\pi^{Aa})$.

Using the Fedosov connection,
we can define  on the superspace ${\cal M}$
the differential operators
corresponding to the covariant derivative on ${\cal M}_0$:
\beq
\nabla_i=\pd_i +\Gamma^k_{ij}\theta_{ka}\frac{\pd}{\pd\theta_{ja}},\qquad
[\nabla_i,\nabla_j]=R^k_{\;mij}\theta_{ka}\frac{\pd}{\pd\theta_{ma}},
\label{nabla}
\eeq
where $R^k_{\;mij}$ are the components of the curvature
tensor~(\ref{R}).

To get a realization of the triplectic algebra in general coordinates,
one can consider a minimal generalization by replacing in the
antibrackets (\ref{trantibr}) the  usual derivative, $\pd_i$, by the
covariant one, $\nabla_i$, namely
\beq
\label{antib}
(F,G)^a=(\nabla_iF)\frac{\pd G}{\pd\theta_{ia}}+(-1)^{(\epsilon(F)}
 \frac{\pd F}{\pd\theta_{ia}}(\nabla_iG)
\eeq
and in the definition of the operators $\Delta^a$,
\beq
\label{Da1}
\Delta^a=\nabla_i\frac{\pd}{\pd\theta_{ia}}
+\hbox{\large$\frac{1}{2}$}(\rho(x),\;\cdot\;)^a~.
\eeq
On the superspace ${\cal M}$ there exists a naturally
defined object being a $Sp(2)$ irreducible second rank tensor, namely
$S_{ab}=(1/6)
\theta_{ia}\omega^{ij}(x)\theta_{ab}$, $S_{ab}=S_{ba}$~.
This tensor is covariantly constant $\nabla_i S_{ab}=0$.
With the help of this tensor we introduce the vector
fields ${V}^a$ in an explicitly  $Sp(2)$ symmetric way
 $V_a=(S_{ab},\;\cdot\;)^b
=(1/2)
\theta_{ia}\omega^{ij}\nabla_j$.

For these objects to be antibrackets,
the operations (\ref{antib}) must satisfy the Jacobi
identities (\ref{Ja}). This requirement has the consequence for
$\Gamma^k_{ij}$ to be a flat connection, $R^k_{\;ijm}=0$.
The existence of such
flat connections directly follows from the Darboux theorem. Namely,
in Darboux coordinates one can choose the trivial connection which is
obviously flat. Note that every non-linear  canonical transformation
transforms the trivial connection into a non-zero one. Such flat connection,
respecting the Poisson bracket, was used in Ref.~\cite{BT} for the
formulation of
a coordinate-free scheme of deformation quantization.

Therefore {\it we arrive at an explicit realization of the triplectic
algebra on ${\cal M}$ with an arbitrary flat Poisson space
${\cal M}_0$}.

To find an explicit realization of the modified triplectic algebra we need
an additional structure.
Let us equip  ${\cal M}_0$
by a symmetric tensor $g_{ij}(x)=g_{ji}(x)$. When this tensor is
nondegenerate it can be considered as a metric tensor on ${\cal M}_0$.
Now we can construct the $Sp(2)$ scalar function $S_0$
which is defined on the superspace $\cal M$ as follows:
$S_0=(1/2)\theta_{ia}g^{ij}\theta_{jb}\vep^{ab}$,
$\epsilon(S_0)=0$, $g^{ij}=\omega^{im}g_{mn}\omega^{nj}$.
It  generates the vector fields $U^a=(S_0,\;)^a$.
Requiring the conditions (\ref{mal}) yields the following equations
for $S_0$
\beq
(S_0,S_0)^a=0 \quad V^aS_0=0,\quad \Delta^a S_0=0.
\label{S1}
\eeq

Let us define the
traceless matrix $I^i_k$ by the relations
$I^i_k=\omega^{ij}g_{jk}$.
In terms of the operator $I$ the relations (\ref{S1}) read
\beq
\nonumber
I^k_{j;i}-I^k_{i;j}=0,\qquad
N^k_{ij}\equiv I^l_iI^k_{j;l} -I^l_jI^k_{i;l}-
I^k_l(I^l_{j;i}-I^l_{i;j})=0,
\eeq
where  $N^k_{ij}$ is the Nijenhuis tensor. When these equations are
fulfilled the set of operators $\Delta^a$, $V^a$
and $U^a$ form the modified triplectic algebra (\ref{malg}),
(\ref{mal}).

\section{Quantization}

The superspace ${\cal M}$
can be equipped with an {\it even symplectic structure}.
Indeed, we can construct an even closed two-form,
which is non-degenerated on the antifields (compare Ref.~\cite{KN}):
\beq
\nonumber
\Omega_2=d(\theta_{ia}\omega^{ij} D\theta^a_j)=
\frac 12 R^{.}_{ijkl}\theta^{ak}\theta^l_a dx^i\wedge dx^j
+\omega^{ij}D\theta_{ia}\wedge D\theta^a_j~,
\eeq
where $D\theta_{ia}=d\theta_{ia}-\Gamma^k_{ij}\theta_{ka}dx^j$,
and $\Gamma^k_{ij}$  is some Fedosov connection (not necessarily flat).
Requiring the connection to be flat,
we can equip the superspace ${\cal M}$,
in addition to the triplectic algebra,
by  the  even symplectic structure
\beq
\nonumber
%\label{ss}
\Omega=dz^\mu\Omega_{\mu\nu}dz^\nu=\omega_{ij}dx^i\wedge dx^j+
\kappa^{-1}\Omega_2=\omega_{ij}dx^i\wedge dx^j+
\kappa^{-1}\omega^{ij}D\theta_{ia}\wedge D\theta^a_j~,
\eeq
where $\kappa$ is an arbitrary constant.
Using this even symplectic structure we can define on
${\cal M}$ the  analogue of the Liouville measure \cite{KGS}
${\cal D}_0={\sqrt{{\rm Ber}\;\Omega_{\mu\nu}}}=
\kappa\left({{\det\;\omega_{ij}}}\right)^{3/2}$.

Let us
construct the vacuum functional and the generating
master equations for a quantum action $S=S(z)$  and a gauge
fixing functional
$X=X(z,\lambda)$ using the basic operators $\Delta^a, V^a, U^a$ and the
function
$S_0$  introduced above. We define the vacuum
functional $Z$ as the following path integral,
\begin{eqnarray}
\label{Z6}
%\nonumber
Z=\int dz \; d\lambda  \;{\cal D}_0\;{\rm exp}
{\{(i/\hbar)[S+X+\al S_0]\}}
\end{eqnarray}
where ${\cal D}_0$ is the integration measure and $\alpha$ is
a constant. We suggest the following form
of generating master equations for $S$ and $X$
\begin{eqnarray}
\label{MEW}
\frac 12(S,S)^a +{\cal V}^aS=i\hbar\Delta^a S,\quad
\frac 12(X,X)^a +{\cal U}^aX=i\hbar\Delta^a X
\end{eqnarray}
where ${\cal V}^a,{\cal U}^a$ are first order differential operators
constructed from $V^a,U^a$. Let us consider the more general transformations
of coordinates $z$,
$ \delta z=(z,F)^a\mu_a + \beta\mu_aV^az +
\gamma\mu_aU^a z$,
generated by the antibrackets $(\;,\;)^a$ and operators
$V^a, U^a$,
where $\beta,\gamma$ are some constants, $\mu_a$ is a $Sp(2)$
doublet of anticommuting constant parameters and  $F=F(z,\lambda)$ is a
bosonic functional.
Choosing $F=X-S$ and
\begin{eqnarray}
\label{VU}
{\cal V}^a=\frac 12(\al U^a+\beta V^a+\gamma U^a),\quad
{\cal U}^a=\frac 12(\al U^a -\beta V^a-\gamma U^a),
\end{eqnarray}
we obtain the invariance of vacuum functional (\ref{Z6})
under the  BRST-antiBRST transformations defined by the
generators $\delta^a=(X-S,\;)^a+{\cal V}^a-{\cal U}^a$.

Evidently
for arbitrary constants $\al, \beta, \gamma$ the operators
${\cal V}^a,\;{\cal U}^a$ obey the properties
${\cal V}^{\{a}{\cal V}^{b\}}=0$,
${\cal U}^{\{a}{\cal U}^{b\}}=0$,
${\cal V}^{\{a}{\cal U}^{b\}}+{\cal U}^{\{a}{\cal V}^{b\}}=0$.
Therefore the operators $\Delta^a$ (\ref{Da1})and ${\cal V}^a,\;{\cal U}^a$
(\ref{VU}) realize the modified triplectic algebra.

That construction, which we derived here, includes all the canonical $Sp(2)$
covariant quantization schemes listed in Section 2:
\begin{itemize}
\item  $\al=0$:
The operators ${\cal V}^a, \;{\cal U}^a$ are linear
dependent ${\cal U}^a =-{\cal V}^a$. In this case there exists only one set
of the vector fields ${\cal V}^a$ and the modified
triplectic algebra is reduced to the triplectic one.

\item
$\al=0,\beta=2,\gamma=0$:
In this case one obtains ${\cal V}^a= V^a$. In fact, this is
representation of vector fields which was used in constructing the
triplectic quantization in general coordinates \cite{BMS}.  In Darboux
coordinates the vector fields ${\cal V}^a$, ${\cal U}^a =-{\cal V}^a$
coincide with the ones used in original version of the triplectic
quantization \cite{BM}.

\item
$\al=1,\beta=2,\gamma=0$:
In the Darboux coordinates  we reproduce the vector fields
${\cal V}^a= V^a$,
${\cal U}^a=U^a$ being used in \cite{BLT}
as well as in the modified triplectic quantization \cite{GGL}.
In this case $\rho=const$,
$S_0=\phi^*_{Aa}\pi^{Aa}$ and the vacuum functional (\ref{Z6}) coincides
with one within the modified triplectic quantization.

\end{itemize}

\section{Conclusions}

We developed (see~\cite{abp}) the extended BRST invariant quantization
scheme in Lagrangian formalism which includes all the ingredients
appeared in the realization of the modified triplectic algebra. The
formulation exactly uses the integration measure extracted from the
even symplectic structure on the whole space of fields and antifields.
It allows   to consider all existing covariant quantization schemes
with extended BRST symmetry  as special limits.
\\

\noindent
\textbf{Acknowledgements.}
P.L. and A.N. acknowledge the hospitality of NTZ at the Center of
Advanced Study of Leipzig University and financial support
by the Saxonian  Ministry of Fine Art and Sciences.
The  work of P.L. was also supported under the INTAS, grant 99-0590
and the projects of Deutsche Forschunsgemeinschaft, 436 RUS 113/669
and Russian Foundation for Basic Research, 02-02-040022.  The work of
A.N.  was supported under  the INTAS project
00-0262 and the ANSEE PS124-01 grant.


\begin{thebibliography}{99}
\bibitem{BV}
I.A.~Batalin and G.A.~Vilkovisky,
\textit{Phys. Lett.}
\textbf{102B} (1981) 27;
\textit{Phys. Rev.} \textbf{28D}
(1983) 2567 [E:  {\bf D30} (1984)
508]; \textit{Nucl. Phys.}
\textbf{234B} (1984) 106.

\bibitem{BLT}
I.A. Batalin, P.M. Lavrov and I.V. Tyutin, \textit{J. Math. Phys.}
\textbf{31} (1990) 1487; {\it ibid.} {\bf 32} (1990) 532;
{\it ibid.}
{\bf 32} (1990) 2513;\\
P.M.~Lavrov, \textit{Mod. Phys. Lett.}
\textbf{6A} (1991) 2051.

\bibitem{BM}
I.A.~Batalin and R.~Marnelius, \textit{Phys. Lett.} \textbf{350B}
(1995) 44.

\bibitem{BMS} I.A.~Batalin, R.~Marnelius and
A.M.~Semikhatov, \textit{Nucl. Phys.} \textbf{446B} (1995), 249;
I.A.~Batalin and R.~Marnelius,
\textit{Nucl. Phys.} \textbf{465B}
(1996) 521.

\bibitem{BM1}
M.~Henneaux, \textit{Phys. Lett.}
\textbf{282B} (1992) 372;  \\ P.~Gregoire and M.~Henneaux,
\textit{Phys. Lett.} \textbf{277B} (1992) 459; \textit{Commun. Math.
Phys.} \textbf{157} (1993) 279.

\bibitem{ND}  A.~Nersessian and P.H.~Damgaard, \textit{Phys. Lett.}
\textbf{355B} (1995) 150;\\
M.A.~Grigoriev and A.M.~Semikhatov,
\textit{Phys. Lett.}
\textbf{417B} (1998) 259;\\
M.A.~Grigoriev, \textit{Phys. Lett.} \textbf{458B} (1999) 499.

\bibitem{GGL}
B.~Geyer, D.M.~Gitman and P.M.~Lavrov,
\textit{Mod. Phys. Lett.} \textbf{14A} (1999) 661;
\textit{Theor. Mat.  Fiz.} \textbf{123} (2000) 476.

\bibitem{W}
E.~Witten, \textit{Mod. Phys. Lett.} \textbf{5A} (1990) 487;

\bibitem{S}
A.~Schwarz, \textit{Commun. Math. Phys.} \textbf{155} (1993) 249;
{\it ibid.} {\bf 158} (1993) 373.

\bibitem{bvgeom} O.M.~Khudaverdian, \textit{J. Math. Phys.}
\textbf{32} (1991), 1934; \\
O.M.~Khudaverdian and A.P.~Nersessian,
 \textit{Mod. Phys.  Lett.} \textbf{8A} (1993) 2377; \textit{J. Math.
Phys.} \textbf{37} (1996) 3713;
\newline H.~Hata and B.~Zwiebach, \textit{Ann.   Phys.(N.Y.)}
\textbf{322} (1994) 131.

\bibitem{F}
B.V. Fedosov, \textit{J. Diff. Geom.} \textbf{40} (1994) 213.

\bibitem{fm}
I.~Gelfand, V.~Retakh and M.~Shubin, \textit{Advan. Math.}
\textbf{136} (1998) 104.

\bibitem{BT}
I.~A.~Batalin and I.~V.~Tyutin,
\textit{Nucl. Phys.} \textbf{345B} (1990) 645.

\bibitem{KN}A.~P.~Nersessian, \textit{JETP Lett.} \textbf{58}
(1993) 66; \textit{Lecture Notes in Physics} \textbf{524}
(1997) 90.

\bibitem{KGS}
O.~M.~Khudaverdian, A.~S.~Schwarz and Y.~S.~Tyupkin,
\textit{Lett. Math. Phys.} \textbf{5} (1981) 517.

\bibitem{abp} B.~Geyer, P.~Lavrov and A.~Nersessian,
\textit{Phys. Lett.} \textbf{512B} (2001) 211;
\textit{Int. J. Mod. Phys.} \textbf{17A} (2002) 1183.
\end{thebibliography}
\end{document}